\documentclass[aps,pre,twocolumn,floatfix,superscriptaddress,showpacs,showkeys]{revtex4-1}
\usepackage{epsf,amsmath,amssymb,verbatim,color,multirow}
\usepackage{graphicx}

\newcommand{\kmin}{{k_{\rm min}}}
\newcommand{\kmax}{{k_{\rm max}}}
\newcommand{\Wsat}{{W_{\rm sat}}}
\newcommand{\WWsat}{{W^2_{\rm sat}}}

\begin{document}
\title{Relevance of the minimum degree to dynamic fluctuation in strongly heterogeneous networks}
\author{H.-H.~\surname{Yoo}}
\author{D.-S.~\surname{Lee}}
\email{deoksun.lee@inha.ac.kr}
\affiliation{Department of Physics, Inha University, Incheon 22212, Korea}
\date{\today}

\begin{abstract}
The fluctuation of dynamic variables in complex networks is known to depend on the dimension and the heterogeneity of the substrate networks. Previous studies, however, have reported inconsistent results for the scaling behavior of fluctuation in strongly heterogeneous networks. To understand the origin of this conflict, we study the dynamic fluctuation on scale-free networks with a common small degree exponent but different mean degrees and minimum degrees constructed by using the configuration model and the static model. It turns out  that the global fluctuation of dynamic variables diverges algebraically and logarithmically with the system size when the minimum degree is one and two, respectively.   Such different global fluctuations are traced back to different, linear and sub-linear, growth of local fluctuation at individual nodes with their degrees,  implying a crucial role of degree-one nodes in controlling correlation between  distinct hubs.
\end{abstract}


\maketitle

\section{Introduction}
\label{sec:intro}

The collective properties of dynamic variables interacting on heterogeneous networks have been the subject of a vast amount of research for their wide applications to various real-world systems~\cite{Watts:1998qf,albert02,barrat2008dynamical} and their importance in the statistical physics of disordered systems~\cite{barabasiBook95,ben2000diffusion,dorogovtsev08}. Among others, the ability to synchronize the activities and balance the loads of individual elements lies in the core mechanism enabling the stable functioning of complex systems and has been investigated intensively in the context of brain networks~\cite{Bullmore:2009aa},  parallel computing~\cite{Korniss31012003}, power transmission~\cite{PhysRevE.69.025103,PhysRevLett.109.064101,Motter:2013aa}, and so on.

Of major importance is understanding how network structure affects the dynamic stability and synchronization~\cite{Gao:2016aa}.  Adopting the Family model~\cite{0305-4470-19-8-006} for the dynamics of load-balancing between neighboring nodes, researchers found that  the fluctuation of loads on scale-free(SF) networks, displaying a power-law degree distribution $P_{\rm deg}(k)\sim k^{-\gamma}$~\cite{barabasi99}, grows logarithmically with the system size and saturates for extremely large system size if the degree exponent $\gamma$ is smaller than 3 and becomes constant even for moderately large system size if $\gamma>3$~\cite{PhysRevE.76.046117,PhysRevE.77.046120,0295-5075-110-6-66001}. Such crucial dependence of dynamic fluctuation on the degree exponent was corroborated in the study of the same dynamics model on SF networks with different ratios of the total number of links to that of nodes~\cite{PhysRevE.93.032319}, in which the fluctuation is shown to diverge with the system size only for $\gamma<3$ if there are sufficiently many links. 

The divergence of dynamic fluctuation for $\gamma<3$ is, however, different between Refs~\cite{PhysRevE.76.046117,PhysRevE.77.046120,0295-5075-110-6-66001} and Ref~\cite{PhysRevE.93.032319}; it grows at most logarithmically with the system size in the former while much faster - algebraically in the latter. This difference may be related to using different model networks, the configuration model network~\cite{Molloy:1995aa,PhysRevE.71.027103}  in the former and the static model~\cite{goh01} in the latter. While the two network models  were used to generate SF networks of similar small degree exponents $\gamma<3$, the structure of the obtained networks are different in many aspects other than the degree exponent. To see which structural factor makes difference in dynamic fluctuation on SF networks, we here use the two network models to obtain SF networks of the same degree exponent but with different minimum degrees and different total numbers of links, and measure the dynamic fluctuation in the Family model dynamics on them.  We find that the scaling behavior of the fluctuation is crucially dependent on the minimum degree of the substrate networks, reminiscent of the Laplacian spectra varying with the minimum degree~\cite{PhysRevE.77.036115}.

We further examine local fluctuations at individual nodes, which grows linearly and sub-linearly with node degree in the networks of the minimum degree one and two, respectively. Considering the linear scaling of fluctuation with the system size in a star graph, we find this result implying that the local fluctuations around distinct hubs are effectively separated only in the networks of the minimum degree one. The scaling exponents characterizing the divergence of the global fluctuation are  reproduced by using the degree dependence of local fluctuation  together with the scaling property of the maximum degree. Given much focus on  hubs in many studies of complex networks, our findings evoke the importance of the low-degree nodes in network structure and dynamics. 

\section{Model}

\subsection{Three groups of substrate networks from the configuration model and the static model}
\label{sec:networkmodel}

To understand the reason why dynamic fluctuation is different between the configuration-model networks and the static-model networks of similar small degree exponents, let us first examine  the properties of the two model networks. 

In the configuration model~\cite{Molloy:1995aa,PhysRevE.71.027103},   each node is first assigned its degree, a random integer $k$ drawn from a given degree distribution $P_{\rm deg}(k)$ between the minimum $\kmin$ and the maximum $\kmax$.    Then randomly selected pairs of link stubs are connected until no stub is left disconnected. To obtain a SF network of $N$ nodes with the degree exponent  $\gamma$, we use the following degree distribution
\begin{equation}
P_{\rm deg}(k) = {k^{-\gamma} \over \sum_{k'=\kmin}^\kmax k'^{-\gamma}} \ {\rm for} \ \kmin\leq k\leq \kmax
\label{eq:Pdeg_config}
\end{equation}
to generate the degree sequence $k_1, k_2, \ldots, k_N$ and connect the link stubs. The obtained network then has the degree distribution as in Eq.~(\ref{eq:Pdeg_config}). Following the above procedures  in case of $\gamma<3$, however, multiple links could be assigned to pairs of nodes having large degrees, which should be rewired to other disconnected nodes, generating negative degree-degree correlation ~\cite{PhysRevE.68.026112}. To avoid generating such degree-degree correlation for $\gamma<3$,   the maximum degree is often restricted to $\kmax = \kmin N^{1/2}$  in the configuration model~\cite{PhysRevE.71.027103}.

In the static model~\cite{goh01}, $L$ links are assigned one by one to each pair of nodes $i$ and $j$ selected with probability $\propto i^{-{1\over \gamma-1}} j^{-{1\over \gamma-1}}$ with $i,j=1,2,\ldots, N$ the node indices. Multiple links are disallowed, which  generates degree-degree correlation for $\gamma<3$ in the static model~\cite{refId0}. The obtained SF networks have the degree distribution behaving  as
\begin{equation}
P_{\rm deg}(k) \simeq (\gamma-1) \left[{2L\over N}\left({\gamma-2 \over \gamma-1}\right)\right]^{\gamma-1}
 k^{-\gamma} 
\label{eq:Pdeg_static}
\end{equation}
for $1\ll k\ll \langle \kmax\rangle$ 
and decaying much faster  for $k\gg \langle \kmax\rangle$ as shown in Eq. (49) of Ref.~\cite{lee04}.  The average  maximum degree is given by $\langle\kmax\rangle = {2L\over N} {\gamma-2\over \gamma-1} N^{1/(\gamma-1)}$~\cite{lee04}. 

The degree exponent $\gamma$ and the number of nodes $N$ are control parameters in both models. The minimum degree $\kmin$ is a parameter only in the configuration model and the total number of links $L$ is a parameter only in the static model. $\kmin$ is most likely to be $1$ unless the ratio $L/N$ is not too large in the static model.  In the configuration model, $L/N$ depends on both $\kmin$ and $\gamma$ via the relation $2L/N = \sum_k k P_{\rm deg}(k)$ with $P_{\rm deg} (k)$ in Eq.~(\ref{eq:Pdeg_config}). 

In Refs.~\cite{PhysRevE.76.046117,PhysRevE.77.046120}, the configuration model with $\gamma=2.5$ and $\kmin=2$ is used to obtain SF networks, which have the mean degree  $\langle k\rangle=2L/N\simeq 4$. In Ref.~\cite{PhysRevE.93.032319}, the static model with $\gamma=2.4$ and $1/4\leq \langle k\rangle\leq 2$  is first used to generate SF networks and their largest-connected components (LCC) are selected for the substrate for running model dynamics since we are interested in  the fluctuation of dynamic variables all pairs of which can interact in principle via a connecting path; Variables in distinct connected components are just independent of one another.  The LCCs  of $N_G<N$ nodes and $L_G<L$ links have the mean degree $\langle k\rangle_G = 2L_G/N_G$ ranging between 2 and 3.2 in Ref.~\cite{PhysRevE.93.032319}. Quantities with subscript $G$ are the properties of the LCC.  
Therefore we see that the substrate networks in the two previous studies have similar low degree exponents but  are different in  the mean degree,  the minimum degree, the maximum degree, and possibly  more properties that remain to be identified. These structural differences can bring different divergence of  dynamic fluctuation despite almost the same degree exponents.

%%%%%%%%%%Table I %%%%%%%%%%%%%%%%
\begin{table}
\centering
\begin{tabular}{c|c|c|c|c|c}
\multirow{2}{*}{group} & \multirow{2}{*}{property} & \multicolumn{4}{|c}{$N$}\\
\cline{3-6}
&&$10^2$ & $10^3$ & $10^4$ & $10^5$\\
\hline
\multirow{4}{*}{(i)} 
& $\langle N_G\rangle/N$ &0.383 & 0.579 & 0.635 & 0.664 \\\cline{2-6}
& $\kmin$ & 1&1&1&1\\\cline{2-6}
& $\langle k\rangle_G$ & 2.03 & 2.27 & 2.44 & 2.56 \\\cline{2-6}
& $\langle \kmax\rangle$ & 7.97 & 27.2& 92.2&306 \\
\hline
\multirow{4}{*}{(ii)} 
& $N_G/N$ &1 & 1 & 1 & 1 \\\cline{2-6}
& $\kmin$ & 2&2&2&2\\\cline{2-6}
& $\langle k\rangle_G$ & 3.69 & 4.28 & 4.71 & 5.00 \\\cline{2-6}
& $\langle \kmax\rangle$ &17.1 &56.7 &189 & 613\\
\hline
\multirow{4}{*}{(iii)} 
& $\langle N_G\rangle/N$ & 0.918& 0.877 & 0.855 & 0.845  \\\cline{2-6}
& $\kmin$ & 1&1&1&1\\\cline{2-6}
& $\langle k\rangle_G$ &4.35 & 4.55 &4.66 & 4.71  \\\cline{2-6}
& $\langle \kmax\rangle$ &29.6 & 143 & 719 & 3740 
\end{tabular}
\caption{Structural properties of three groups of SF substrate networks. All have the same degree exponent  $\gamma=2.4$.  The number of nodes $N_G$, the minimum degree $\kmin$ and the mean degree $\langle k\rangle_G$ are shown. For group (i) and (iii), the LCCs are used for our study, which may have smaller numbers of nodes $N_G$ than the total number of nodes $N$ of the original networks.}
\label{table:networkmodel}
\end{table}
%%%%%%%%%%%%%%%%%%%%%%%%%%

%%%%%%  Figure 1%%%%%%%%%%%%%%%%%
\begin{figure}
\includegraphics[width=\columnwidth]{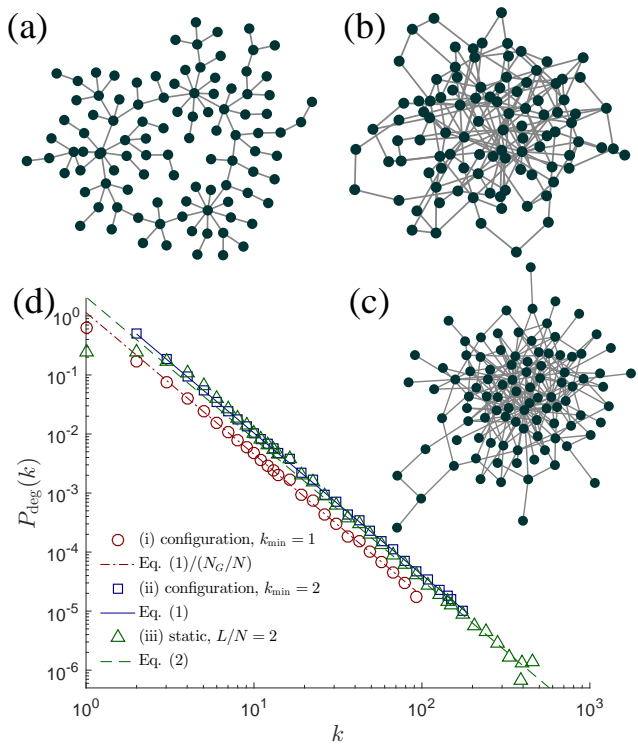}
\caption{Examples of the  SF substrate networks of degree exponent $\gamma=2.4$ belonging to three groups.
(a) A network in group (i) of $N_G=101$ nodes, $L_G=102$ links, the minimum degree $\kmin=1$, and the maximum degree $\kmax=10$.
(b) A network in group (ii) of $N_G=100$, $L_G=193$, $\kmin=2$, and $\kmax=18$.  
(c) A network in group (iii) of $N_G=99$, $L_G=219$, $\kmin=1$, and $\kmax=32$. 
(d) Degree distributions of the three groups of substrate networks of $N=10^4$.  Lines represent theoretical predictions, $P_{\rm deg}(k)/(\langle N_G\rangle/N)$ with $P_{\rm deg}(k)$ in Eq.~(\ref{eq:Pdeg_config})  for group (i),   $P_{\rm deg}(k)$ in Eq.~(\ref{eq:Pdeg_config}) for group (ii), and  $P_{\rm deg}(k)$ in Eq.~(\ref{eq:Pdeg_static}) for group (iii).} 
\label{fig:modelnetworks}
\end{figure}
%%%%%%%%%%%%%%%%%%%%%%%%%%

To pinpoint the  structural factors relevant to dynamic fluctuation on SF networks with  low degree exponents, we consider in this work the following three groups of model networks: \\ \\
(i) the LCCs of the networks of $N$ nodes generated by the configuration model with  $\gamma=2.4$ and $\kmin=1$,\\  
(ii) the connected networks of $N$ nodes generated by the configuration model with $\gamma=2.4$ and $\kmin=2$, and \\
(iii) the LCCs of the networks of $N$ nodes generated by the static model with $\gamma=2.4$ and $L/N=2$.  \\ \\
The substrate networks in group (ii) are similar to those used in Refs.~\cite{PhysRevE.76.046117,PhysRevE.77.046120,0295-5075-110-6-66001} and the substrates in (iii) are similar to those in Ref.~\cite{PhysRevE.93.032319}.  The networks in group (i) are newly considered here for our comparative study. Note that most of the networks generated by the configuration model with $\gamma=2.4$ and $\kmin=2$ are not fragmented but connected, being a single connected component~\cite{PhysRevE.76.046117,PhysRevE.77.046120,0295-5075-110-6-66001}.  For group (ii), we abandon few networks that are fragmented. 

The size of these substrate networks, the LCC's or the connected networks, will be called the substrate size or the system size and denoted by $N_G$, distinguished from the total number of nodes $N$ in the original networks. Some of their structural properties  are presented in Table~\ref{table:networkmodel} and examples are shown in Fig.~\ref{fig:modelnetworks} (a-c).  The networks in group (i) have smaller mean degree than those in (ii) and (iii). The model networks in (i) and (iii)  have the minimum degree one but  group (ii) do not have a node of degree one. Despite these differences, they  share the same asymptotic behavior of the degree distribution  as shown in Fig.~\ref{fig:modelnetworks} (d). 

\subsection{Dynamics: Family model}

On the substrate networks introduced in Sec.~\ref{sec:networkmodel},  we consider the Family model~\cite{0305-4470-19-8-006} for the interaction and time-evolution of  dynamic variables. An integer-valued variable $h_i$ at node $i$, representing its activity, work load  or the number of particles piled at the node, may grow with time under a diffusive force towards reducing the difference of $h$'s between neighboring nodes.  Let us  use the terminology from the viewpoint that $h_i$ is the number of particles piled at node $i$, called height, and $\{h\}$ represents surface heights on a given substrate network. 

The time-evolution of the surface heights in the Family model is given as follows~\cite{0305-4470-19-8-006,barabasiBook95,PhysRevE.76.046117,PhysRevE.77.046120,0295-5075-110-6-66001,PhysRevE.93.032319}.  Initially all heights are zero. 
At each time step,  a node $i$ is randomly selected. If  $h_i$ is not larger than any of the neighbor nodes' heights, i.e., $h_i\leq h_j$  for all $j\in n.n.(i)$, then $h_i$ is increased by one, $h_i \to h_i +1$. If $h_i$ is larger than at least one neighbor's height, then  the lowest of the neighbors' heights is  increased  by one, that is, $h_\ell \to h_\ell +1$ with    $h_\ell = \min_{j\in n.n.(i)}h_j $.  These procedures are repeated every time step. 

We are mainly interested in how much $h$'s are different from one another. To quantify it, we compute the fluctuation $W(t)$ of $h$'s, which is called the roughness  in the context of surface growth and evaluated for a substrate of $N_G$ sites  as 
\begin{equation}
W(t)=\left\langle {1\over N_G} \sum_{i=1}^{N_G} (h_i(t)-\bar{h}(t))^2\right\rangle^{1/2}.
\label{eq:Wdef}
\end{equation}  
Here $\bar{h}(t)$ is the spatial average of heights $\bar{h}(t) = \sum_{i=1}^{N_G} h_i(t)/N_G$ and $\langle \cdots \rangle$ indicates the average over different realizations of the dynamics and of the topology of substrates.

\section{Divergence of global fluctuation}

%%%%%% Figure 2 %%%%%%%%%%%%%%%%%%
\begin{figure}
\includegraphics[width=\columnwidth]{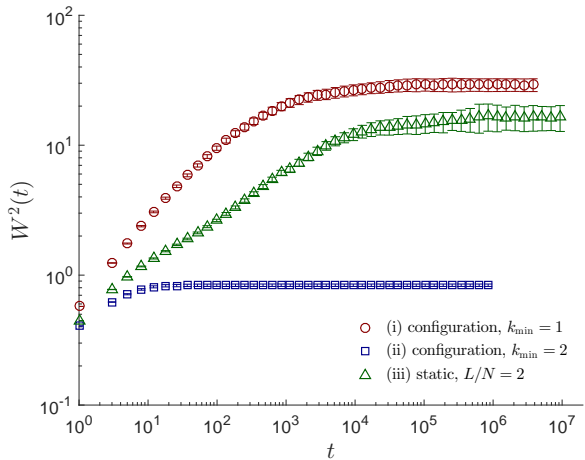}
\caption{Time evolution of the dynamic fluctuation (roughness) $W^2(t)$  in the Family model on the SF networks with the degree exponent $\gamma=2.4$ and  $N=10^5$ in three groups (i), (ii), and (iii). Average over 100 different realizations of networks and dynamics is taken for each case. }
\label{fig:W2vst}
\end{figure}
%%%%%%%%%%%%%%%%%%%%%%%%%%%%%%

The variations of the roughness $W(t)$  with time $t$ and the substrate size $N_G$ have been of great interest in the study of surface roughening~\cite{barabasiBook95}. In general the roughness initially increases with time as  identified also in our simulations [Fig.~\ref{fig:W2vst} (a)].  The roughness  saturates in the long-time limit for finite system size.  The saturated roughness $W_{\rm sat}$ is measured by taking the time average of the simulation results in the stationary state.  For many surface growth models, including the Family model, the saturated roughness on the Euclidean lattice of size $N_G$ displays a scaling behavior~\cite{barabasiBook95} 
\begin{equation}
W_{\rm sat} \sim N_G^\alpha.
\label{eq:scaling}
\end{equation}
The larger the scaling exponent $\alpha$ is,  the larger the fluctuation of the dynamic variables is. If the scaling exponent $\alpha$ is zero, the roughness remains finite even in the thermodynamic limit $N_G\to\infty$, meaning that the dynamic variables $h$'s show negligible fluctuation. If $\alpha>0$, the roughness diverges for infinite system size. It has been known that the exponent depends on the dimension of the substrate lattice and the fundamental properties of the growth model, such as whether the corresponding Langevin equation includes a non-linear term or not~\cite{barabasiBook95}.

%%%%%% Figure 3 %%%%%%%%%%%%%%%%%%
\begin{figure}
\includegraphics[width=\columnwidth]{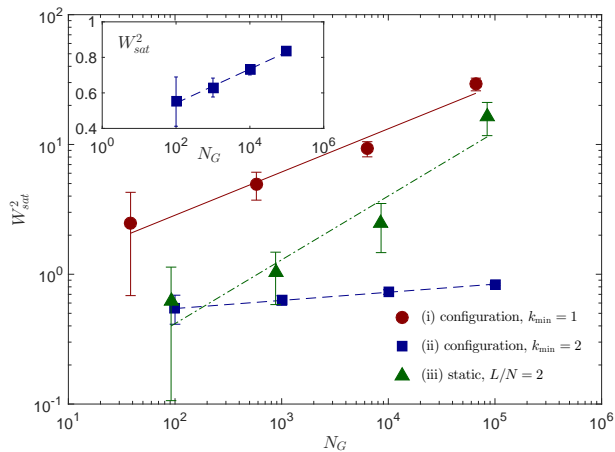}
\caption{Saturated roughness $W_{\rm sat}^2$ versus the system size $N_G$ plotted in logarithmic scales. The fitting lines have slope $2\alpha$ with $\alpha =0.167$, $0.0319$, and $0.246$ for group (i), (ii), and (iii), respectively.  The inset shows the  data for group (ii) plotted in semi-logarithmic scale. The line is fitted to data,  $\WWsat =0.0417 \log N_G + 0.351$. }
\label{fig:W2satNG}
\end{figure}
%%%%%%%%%%%%%%%%%%%%%%%%%%%%%%

For the Family model on our model SF networks, the saturated roughness also grows with the substrate size $N_G$ as shown in Fig.~\ref{fig:W2satNG}. Assuming that Eq.~(\ref{eq:scaling}) is valid, we find that  
\begin{align}
\alpha_{\rm (i)} &= 0.167 \pm 0.030, \nonumber \\
\alpha_{\rm (ii)} &= 0.0319\pm 0.0098, \nonumber \\
\alpha_{\rm (iii)} &= 0.246\pm 0.041,
\label{eq:alpha}
\end{align}
where the subscripts indicate the groups of model networks. Despite having the same degree exponent $\gamma=2.4$,  they have  different values of $\alpha$. 
In particular,  the networks in group (ii), the configuration model with $\kmin=2$, has $\alpha$ close to zero;  actually the saturated roughness scales logarithmically with the system size [Fig.~\ref{fig:W2satNG}], which is consistent with the previous results~\cite{PhysRevE.76.046117,PhysRevE.77.046120}. The groups (i) and (iii), the configuration model networks with $\kmin=1$ and the static model network have positive values of $\alpha$, about 0.17 and 0.25 while they have different mean degrees $\langle k\rangle_G$ around 2 and 4, respectively. In the next two sections, we investigate the origin of such different values of $\alpha$ in  Eq.~({\ref{eq:alpha}).

\section{Spectral dimension}

For the Edwards-Wilkinson (EW) model dynamics~\cite{Edwards17}, a linear diffusion model, the roughness can be represented as $W^2_{\rm sat} = N_G^{-1} \sum_{n=2}^\infty  \lambda_n^{-1}$ with  $\lambda_1=0 <\lambda_2\leq \lambda_3\cdots \leq \lambda_{N_G}$ the eigenvalues of the Laplacian matrix $L_{ij} = k_i \delta_{ij} - A_{ij}$ of the substrate network having adjacency matrix $A_{ij}$~\cite{hwang2014fast}. It has been shown that this exact result can be used  to understand the positive values of the scaling exponent $\alpha$  also in the Family model on  the substrates of the spectral dimension $d_s$ smaller than $2$; If $d_s<2$, then   the roughness exponent is given by $\alpha = {1\over 2} \left({2\over d_s} -1\right)$~\cite{PhysRevE.93.032319}. The spectral dimension $d_s$ characterizes the small-$\lambda$ behavior of the spectral density function $\rho(\lambda)\equiv {1\over N_G} \sum_{n=1}^{N_G} \delta(\lambda-\lambda_n)\sim \lambda^{d_s/2 -1}$ of the Laplacian matrix~\cite{ben2000diffusion}.  Such large fluctuation as diverging with the system size appears for substrates of small spectral dimension, which can be understood naively by considering that fewer pairs of interacting neighbors in the substrates of small $d_s$   than those of large $d_s$ may make difficulty in achieving  global synchronization.  

%%%%%% Figure 4 %%%%%%%%%%%%%%%%%%
\begin{figure}
\includegraphics[width=\columnwidth]{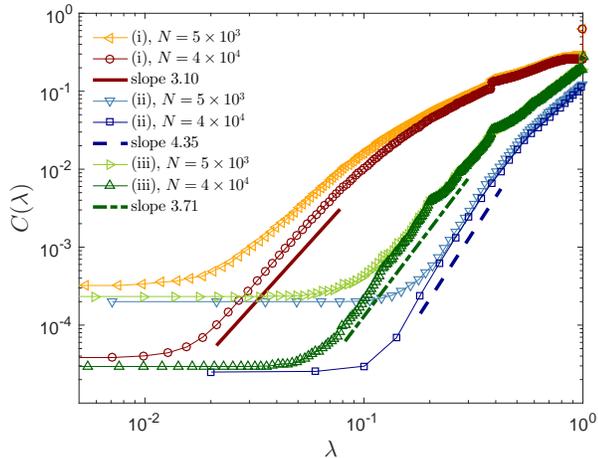}
\caption{Cumulative distribution $C(\lambda)$ of the eigenvalues of the Laplacian matrix of the SF networks for different $N$'s in three groups (i), (ii), and (iii). The lines fitting Eq.~(\ref{eq:ds}) to data for small $\lambda$ are also shown with their slopes given in the legend.}
\label{fig:cSDF}
\end{figure}
%%%%%%%%%%%%%%%%%%%%%%%%%%%%%%

To check the possibility that different values of $\alpha$ in  Eq.~(\ref{eq:alpha}) stem from different dimensionality of the substrate networks, we estimate their spectral dimensions $d_s$.  To do so, we obtain numerically the eigenvalues of the Laplacian matrices of the substrate networks in three groups and compute the cumulative distribution $C(\lambda) = \int_\lambda^\infty d\lambda \, \rho(\lambda)$. The spectral dimension $d_s$ can be estimated by fitting 
\begin{equation}
C(\lambda) \sim \lambda^{d_s\over 2}
\label{eq:ds}
\end{equation}
to data for $\lambda$ small.  In Fig.~\ref{fig:cSDF}, we find that  the estimated spectral dimensions of all  SF networks in three groups are far larger than the critical dimension $2$;   $d_s = 6.20\pm 0.75$, $8.70\pm 1.14$, and $7.42\pm 0.64$ for model (i), (ii), and (iii), respectively. Small fluctuation characterized by the low value of $\alpha$ in the group-(ii) networks, obtained by the configuration model networks with $\kmin=2$, may be attributed  to such high value of $d_s$. On the contrary,  the diverging fluctuations in group-(i) and -(iii) networks must have origins other than the dimensionality of the substrates.

The roughness is affected not only by the dimensionality but also by the connection heterogeneity of the substrates~\cite{PhysRevE.93.032319}, which can be a key to understanding  the diverging fluctuation in the SF networks in group (i) and (iii) having large $d_s$; heterogeneous networks have a non-negligible fraction of hubs, which can induce globally large fluctuation. Whether  heterogeneity in connectivity can bring such diverging fluctuation or not depends on the degree-dependent behavior of the local fluctuation at individual nodes, which is investigated in detail for our model networks in the next section. 

\section{Local fluctuation}

Since the locally-lowest height is increased by one every time step in the Family model, the deviation of the height at node $i$ from the spatial average height $\Delta h_i(t) =  h_i(t) - \bar{h}(t)$ may depend on the degree $k_i$ as well as  the specific realization of the dynamics. Assuming that the degree distribution does not vary significantly with the realization of the substrate networks for a given group considered in this work, one can decompose the global fluctuation into local fluctuations according to node degree as 
\begin{equation}
W^2(t) = \left \langle {1\over N_G} \sum_{i=1}^{N_G} [\Delta h_i(t)]^2 \right\rangle  = \sum_k P_{\rm deg} (k) W^2(k,t),
\label{eq:W2decompose}
 \end{equation}
where $W^2(k,t)$ is  the local fluctuation  at nodes of degree $k$ and defined as 
\begin{equation}
W^2(k,t) = \left\langle {\sum_{i=1}^{N_G} \delta_{k_i,k} [\Delta h_i(t)]^2\over \sum_{i=1}^{N_G} \delta_{k_i,k}} \right\rangle.
\end{equation}
Measuring the degree-dependent local fluctuation $\WWsat(k)$ and the degree distribution $P_{\rm deg}(k)$ in simulations and inserting them into Eq.~(\ref{eq:W2decompose}),  one can evaluate the global fluctuation.

%%%%%% Figure 5 %%%%%%%%%%%%%%%%%%
\begin{figure}
\includegraphics[width=\columnwidth]{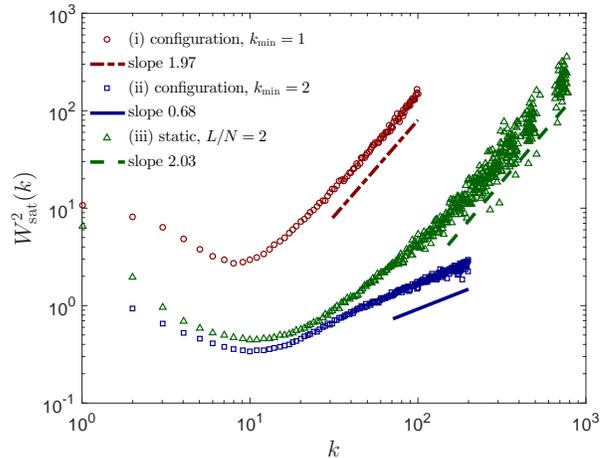}
\caption{Local fluctuation $W^2(k)$ at nodes of degree $k$ in the Family model on the SF networks in three groups. The lines fit Eq.~(\ref{eq:W2ktheta}) to data for large $k$ and  their slopes $2\theta$ are given in the legend.  }
\label{fig:W2k}
\end{figure}
%%%%%%%%%%%%%%%%%%%%%%%%%%%%%%

Taking the time-average of the simulation data for $W^2(k,t)$ in the stationary state, we obtain the saturated local fluctuation $W^2_{\rm sat}(k)$, which is found to behave as 
\begin{equation}
W_{\rm sat}(k) \sim k^\theta
\label{eq:W2ktheta}
\end{equation}
for large $k$ in all three groups of model networks [Fig.~\ref{fig:W2k}]. The scaling exponent $\theta$ is, however, different:
\begin{align}
\theta_{\rm (i)} &= 0.99\pm 0.03,\nonumber\\
\theta_{\rm (ii)} &= 0.34\pm 0.02,\nonumber\\
\theta_{\rm (iii)} &= 1.02\pm 0.03.
\label{eq:theta}
\end{align}
It is around $1$ for group (i) and (iii), having $\kmin=1$ while $\theta$ is far smaller than $1$ for group (ii) having $\kmin=2$.

Large local fluctuation at nodes of large degree can be understood by considering the height distribution in  the Family model on a star graph of $k+1$ nodes, consisting of one center node connected to $k$ peripheral nodes of degree one. The center node is more frequently selected to increase its height than peripheral nodes, resulting in its height higher than the spatial average height. Simultaneously the heights at peripheral nodes are made very different from one another. Simulations show that  the local fluctuation at the center node scales linearly with the graph size $k$, $W_{{\rm sat},0} \sim k$ for large $k$~\cite{PhysRevE.93.032319}. 

In SF networks, a subgraph consisting of a hub node of large degree $k$ and its $k$ neighbors may be viewed as a star graph. If the dynamic variables in distinct such star-like subgraphs are independent, then one can expect that the local fluctuation $W_{\rm sat}(k)$ will scale linearly with $k$ like  the star graph of $k+1$ nodes.  It is indeed the case for group (i) and (iii) as given in Eq.~(\ref{eq:theta}). Abundant degree-one nodes are expected to enable the subgraphs formed around distinct hubs to be dynamically independent.  
On the contrary,  the local fluctuation shows a sub-linear scaling with degree for SF networks in group (ii), which implies that the star subgraphs around distinct hubs are correlated.  We conjecture that enforcing the minimum degree  $2$ in group-(ii) networks may enhance such inter-dependence between distinct hub neighborhoods. 

The degree-dependent behavior of local fluctuation allows one to derive the scaling exponent $\alpha$. Inserting $P_{\rm deg}(k)\sim k^{-\gamma}$ and  the $W^2_{\rm sat}(k)$ of Eq.~(\ref{eq:W2ktheta})  into Eq.~(\ref{eq:W2decompose}), we find that 
\begin{align}
\WWsat \sim \sum_{k\gg 1}^{\kmax} k^{-\gamma} k^{2\theta} \sim \left\{
\begin{array}{ll}
\kmax^{1-\gamma+2\theta} & \ {\rm for} \ 1-\gamma+2\theta>0, \\
{\rm const.} & \ {\rm for} \ 1-\gamma+2\theta<0.
\end{array}
\right.
\label{eq:W2kmax}
\end{align}
Here we assumed that the scaling behavior of $\WWsat$ is driven by the contributions from the fluctuations at nodes of large degrees as long as the spectral dimension of the substrate is larger than 2; It has been shown that the fluctuations at nodes of small degrees and those of large degrees scale with respect to  $N_G$ in the same way~\cite{PhysRevE.93.032319}. To evaluate Eq.~(\ref{eq:W2kmax}), the $N_G$-dependence of the maximum degree $\kmax$ should be used, which is discussed in Sec.~\ref{sec:networkmodel} and behaves  as 
\begin{equation}
\kmax\sim N_G^\eta
\label{eq:kmax}
\end{equation}
 with the exponent $\eta$ given by 
\begin{align}
\eta_{\rm (i)} &= {1\over 2}, \nonumber\\
\eta_{\rm (ii)} &= {1\over 2},\nonumber\\
\eta_{\rm (iii)} &= {1\over \gamma-1}.
\label{eq:eta}
\end{align} 
Using Eqs.~(\ref{eq:kmax}) and (\ref{eq:eta}) into Eq.~(\ref{eq:W2kmax}), we find that $\Wsat \sim N_G^\alpha$ with the roughness scaling exponent $\alpha$ given by 
\begin{equation}
\alpha = \max\left\{{\eta \over 2} \left(1-\gamma+2\theta\right),0\right\} =\left\{
\begin{array}{ll}
0.15\pm 0.02 & \ {\rm for \ (i),}\\
0  & \ {\rm for \ (ii),} \\
0.23\pm 0.02 & \ {\rm for \ (iii).}
\end{array}
\right.
\end{equation}
These are in good agreement with $\alpha$'s given in Eq.~(\ref{eq:alpha}) estimated directly from the $N_G$ dependent behaviors of $\Wsat$.  Therefore we can conclude that the diverging fluctuation in SF networks of $\gamma=2.4$ and $\kmin=1$ originates in the strong heterogeneity of degrees.  Different values of the exponent $\alpha$ between group (i) and (iii), although having $\kmin=1$ and $\gamma=2.4$ in common, originate in the different scaling behaviors of the maximum degree given in Eq.~(\ref{eq:eta}). For the SF networks in group (ii), the local fluctuation does not grow fast enough with degree  to lead the global fluctuation to diverge algebraically but resulting in $\alpha=0$. 
%This is reasonably consistent with the weak - logarithmic divergence or convergence of the roughness with respect to the substrate size~\cite{0295-5075-110-6-66001}. 

These findings suggest that the minimum degree is a key factor determining the scaling behavior of fluctuation in the SF networks with low degree exponents. While the SF networks in group (ii) and (iii) look similar to some extent [Fig.~\ref{fig:modelnetworks}] due to their similar numbers of links per node, the dynamic fluctuation is almost finite in (ii) but diverging algebraically in (iii). Their structural difference stems from different minimum degrees; Despite similar total numbers of links, a lot of nodes have degree one  in group-(iii) networks but  no node has degree one in group-(ii) networks. Such abundant degree-one nodes in group (i) and (iii) may sustain the inter-dependence between the dynamic variables around distinct hubs, leading to $\theta\simeq 1$ and $\alpha$ positive. On the contrary,  all nodes having at least two links in group (ii) may induce distinct hubs to be connected by multiple paths, thereby enhancing their correlations and reducing dynamic fluctuation.

\section{Discussion}

Here we investigated dynamic fluctuations in three groups of SF networks having the same low degree exponent but  different minimum degrees, mean degrees, and maximum degrees. We found that the SF networks with the minimum degree one may have large fluctuation, diverging algebraically with the system size. On the other hand, the SF networks with the minimum degree two have small fluctuation, at most logarithmically diverging with the system size.

To understand the origin of such strikingly different fluctuations in the SF networks of a common low degree exponent, we examined their spectral dimensions and  local fluctuations. All the studied substrate networks have spectral dimensions larger than 2, negating the possibility that the dynamic fluctuations on them are driven by the low dimensionality.  However, the local fluctuation shows big difference; It grows linearly with node degree in the SF networks of the minimum degree one but sub-linearly in those of the minimum degree two. Such linear or sub-linear behaviors of the local fluctuation enable us to derive analytically the scaling behaviors of the global dynamic fluctuation in the studied SF networks, which are in excellent agreement with the simulation results. It might be  surprising that the minimum degree is so crucial as to differentiate the behavior of the local fluctuation and eventually that of the global fluctuation in SF networks. We argued that the inter-dependence between the dynamic variables around different hubs is crucial for the behavior of local fluctuation and may vary with  the amount of degree-one nodes which can bring different pathway organization.

Our study suggests that the measurement of  the network characteristics relevant to the dynamic fluctuation, that is, the spectral dimension, the degree exponent, and the minimum degree,  can be a first step towards understanding the stability and fluctuation of various complex systems. While we compared the extreme cases of degree-one nodes abundant or absent, it will be of interest to continuously vary the amount of degree-one nodes and see the variation of  local and global dynamics fluctuation and more other  features.  

\acknowledgements
This work was supported by Inha University research grant (No. 53670). We thank Prof. Jin Min Kim for insightful comments. 

%\bibliography{/Users/dslee/Dropbox/Bibliography/Master}


\end{document}